\documentclass[aps,prb,twocolumn,letterpaper]{revtex4}
\usepackage{graphicx}
\usepackage{dcolumn}
\usepackage{amsmath}
\usepackage{amssymb}
\usepackage{float}
\usepackage{epstopdf}
\usepackage{color}

\def\kv{{\bf k}}

\begin{document}

\title{Topological Superconductivity without Proximity Effect}
\author{Aaron Farrell and T. Pereg-Barnea}
\affiliation{Department of Physics and Center for the Physics of Materials, McGill University,
Montreal, QC, Canada}
\date{\today}
\begin{abstract}
Majorana Fermions, strange particles that are their own antiparticles, were predicted in 1937 and have been sought after ever since.  In condensed matter
they are predicted to exist as vortex core or edge excitations in certain exotic superconductors.  These are topological superconductors whose order
parameter phase winds non-trivially in momentum space.  In recent years, a new and promising route for realizing topological superconductors has opened due
to advances in the field of topological insulators.  Current proposals are based on semiconductor heterostructures, where spin-orbit coupled bands are
split by a band gap or Zeeman field and superconductivity is induced by proximity to a conventional superconductor.  Topological superconductivity is
obtained in the interface layer.  The proposed heterostructures typically include two or three layers of different materials.  In the current work we
propose a device based on materials with inherent spin-orbit coupling and an intrinsic tendency for superconductivity, eliminating the need for a separate
superconducting layer.  We study a lattice model that includes spin-orbit coupling as well as on-site and nearest neighbor interaction.  Within this model
we show that topological superconductivity is possible in certain regions of parameter space.  These regions of non-trivial topology can be understood as a
nodeless superconductor with d-wave symmetry which, due to the spin-orbit coupling, acquires an extra phase twist of $2\pi$.
\end{abstract}
\maketitle


\section{Introduction}
Back in 1937, Ettore Majorana found particles that arise as real solutions to the Dirac equation.  These solutions, called Majorana fermions, partially
obey fermionic statistics.  While different Majorana fermions anti-commute, each Majorana fermion is its own anti particle.  The creation operator of a
Majorana fermion is also its annihilation operator or, in other words it is an equal superposition of regular fermionic creation and annihilation
operators.  This exceptional property captured the imagination of many and the quest to find the Majorana fermion began.

In the context of high energy physics it is speculated that the neutrino might in fact be a Majorana fermion. The testing of this claim, which originated
from Majorana himself, has not been possible in the past, as it requires a large collider like the LHC and is still an open question at the time of
writing.

Regardless of the nature of the neutrino and possible other elementary particles of the Majorana type, Majorana fermions may be realized in condensed
matter systems.  In condensed matter, excitations are not limited to elementary particles since they may be emergent particles that are dressed by the
medium and interactions in the many body state.  As such, it is conceivable that emergent excitations be their own anti-particles.  Furthermore, in
condensed matter anti-particles are provided by holes in energy bands and the superposition of particles and holes is possible.  Such a superposition occurs as an excitation in any superconductor, and the number of particles is not conserved due to the presence of a pair
condensate.

In order to realize Majorana fermions, a system should exhibit pairing between two particles of the same spin.  This requires triplet pairing and
particularly a complex $p$-wave order parameter is desirable.  It has been shown\cite{Read} that topological, spin-triplet, $p_x+ip_y$ superconductors will
support Majorana fermions in their vortex cores\cite{Gurarie,Sato2}. Some materials have been found to have triplet $p$-wave pairing; however their topology has
yet to be proven to be non-trivial\cite{Kallin}.  Therefore, current efforts to realize Majorana fermions have had to focus on devices which lead to
quantum states that are either topological superconductors or analogous to them.\cite{Read,Fu,Sau,Alicea,Cook} An interesting analogue of a topological
superconductor was proposed to describe the fractional quantum Hall state in some fractions\cite{Read} and some progress in this direction has been
made\cite{Bonderson,Stern,Dolev}.  In that state, Chern-Simons dressed particles minimize their interaction energy by creating a condensate whose symmetry
is $p_x+ip_y$. Proving beyond doubt the existence of this state as well as the detection of Majorana fermions therein still remains a challenge.

Recently, inspired by advances in topological insulators, another route to topological superconductivity has opened.  Fu and Kane\cite{Fu} have shown that
a three dimensional topological insulator layer placed in proximity to a conventional $s$-wave superconductor develops topological superconductivity. The
pairing in the system is induced by proximity effect while the topology is inherited from the topological insulator.  This occurs since the pairing
function is projected to one of the spin-orbit coupled bands.  In order to accommodate the unique spin structure (and Chern number) of the topological
insulator, the induced order parameter must wind its phase by $2\pi$ in momentum space.

The idea of Fu and Kane was further developed by Tanaka {\it et al.}\cite{Tanaka} who proposed placing junctions containing superconductors on a three dimensional topological insulator.  Sau {\it et al.}\cite{Sau} eliminated the need for a topological insulator and envisioned a semiconductor quantum well with intrinsic Rashba spin-orbit coupling (SOC) where the Fermi surface lies in the band.  The required gap between the two
spin-orbit coupled bands is provided by an out of plane Zeeman field of an attached ferromagnetic (FM) insulator layer.  Meanwhile, superconductivity is
induced by proximity to a superconducting layer attached to the other side of the quantum well.

Recently, Alicea\cite{Alicea} explored the possibility of eliminating the ferromagnetic insulator layer of the Sau {\it et al} model. Instead of the
ferromagnetic insulating layer, Alicea suggested using a quantum well with both Dresselhaus and Rashba SOC while applying an in plane magnetic field. The
Dresselhaus SOC tilts the plane in which the electron spins tend to align so that the applied magnetic field can open a gap, eliminating the need for the
FM insulator and thereby reducing the complexity of the device.

Other suggestion for the realization of Majorana fermions were made in the context of quasi one dimensional structures such as nano-wires and nano-tubes.  Typically these proposals contain strong spin orbit coupling and proximity induced superconductivity\cite{lutchyn,oreg,Cook,klinovaja}.

A general argument relating the type of superconducting order parameter and symmetries of the model was explored by Fu and Berg\cite{FuBerg}.

The key aspects of realizing an effective $p+ip$ state in these previous devices has been the proper combination of SOC, band gap or Zeeman splitting and
proximity induced pairing.  The spin orbit coupling is responsible for the non-trivial spin texture, whereas the Zeeman field (or an intrinsic mass term)
splits the bands such that only one of them is relevant at low energy. Meanwhile, the superconductor responsible for inducing pairing through proximity is
of a simple singlet type.

In this Letter we use these key ingredients to address the question of whether a topological superconductor can be achieved without proximity effect. In
place of proximity induced pairing we consider pairing driven by interactions. A general proof-of-principle that interactions can indeed lead to a topological superconducting state has been provided in Reference [\onlinecite{tewari}]. Using a variational mean field approach, both phases of trivial and
topological superconductivity are found in the model studied. The topological state we find can be described by a superconductor with a $6\pi$ phase
winding which is a result of the $l=2$ d-wave phase winding and a $p+ip$ projection function.


\section{Model and Methods}
In order to test whether interactions may lead to superconductivity in spin-orbit coupled materials we consider a two dimensional
square lattice model.  The Hamiltonian of the system reads
\begin{equation}\label{eq:FullH}
 H= H_{\text{KE}}+H_{\text{SO}}+V,
\end{equation}
where the kinetic energy term $H_{KE}$ is given by hopping on nearest neighbors.
\begin{equation}
H_{\text{KE}} = -t\sum_{\langle i,j \rangle,\sigma} (c^\dagger_{i\sigma}c_{j\sigma}+c^\dagger_{j\sigma}c_{i\sigma}).
\end{equation}
Here $t$ is the hopping amplitude, $\langle i,j \rangle$ are nearest-neighbor lattice sites and $\sigma$ is a spin index. The spin-orbit coupling part of
the Hamiltonian is given by
\begin{equation}\label{SOC}
H_{\text{SO}} = \sum_{{\bf k}} \psi^\dagger_{\bf k} \mathcal{H}_{\bf k} \psi_{\bf k}, \ \ \mathcal{H}_{\bf k} = \sigma \cdot {\bf d}_{\bf k}
\end{equation}
where $\psi_{\bf k} = (c_{{\bf k}\uparrow},c_{{\bf k}\downarrow})^T$, $\sigma$ is a 3-vector of Pauli matrices and $d_{\bf
k}=(A\sin{k_x},A\sin{k_y},2B(\cos{k_x}+\cos{k_y}-2)+M)$ with $A,B$ and $M$ material parameters.  In the above we have assumed units where the lattice constant $a=1$. This term can be viewed as the lattice version of the
continuum model introduced by previous authors\cite{Fu,Sau,Alicea} while its form resembles one of the sectors of the model introduced by Bernevig, Hughes and Zhang (BHZ) to describe HgTe quantum wells\cite{BHZ}.

The three parameters in the spin-orbit coupling model above may originate from a variety of different sources. For example, the parameters $A,B$ may be traditional spin-orbit coupling terms like the Rashba and Dresselhaus terms in Refs. [\onlinecite{Sau}] and [\onlinecite{Alicea}] or may be parameters such as those used in the BHZ model\cite{rothe, lu, guigou}. Similarly, the ``mass" term $M$ may be the result of a band gap\cite{BHZ}, an external magnetic field or a magnetic
field of a nearby ferromagnetic layer\cite{Sau}. As $M$ may come from a variety of sources we will ignore any orbital effects that could arise in the specific case that it comes from a magnetic field. If it does happen that $M$ is from an applied field, we will assume orbital effects to be small. The issue of orbital effects when $M$ arises from a magnetic field are discussed in Reference [\onlinecite{Sato1}].

The reader should note that the versatility of our model for hopping plus spin-orbit coupling, $H_{KE}+H_{SO}$, leads to typical values of the parameters $A,B,M$ stretching over a rather large range. In the case where one is concerned with $A$ and $B$ coming from Rashba like contributions, $A$ and $B$ will be small\cite{Alicea} compared to $t$. There is also the case where $H_{SO}$ is taken to mimic one sector of the model of BHZ. To be more explicit let us recall this model here
\begin{equation}\label{hbhz}
\mathcal{H}_{BHZ} = \left( \begin {array}{cc} \mathcal{M}(k)-\tilde{D}k^2&\tilde{A}k_{-}\\ \noalign{\medskip}\tilde{A}k_{+} &-\mathcal{M}(k)-\tilde{D}k^2\end {array}
\right)
\end{equation}
where $\mathcal{M}(k) = \tilde{M}-\tilde{B}k^2$, $k_{\pm} = k_x\pm i k_y$, $k^2=k_x^2+k_y^2$ and we have used the tilde symbol to differentiate between our model parameters and the ones in the model above. If one discretizes the above model by sending $k_i \to \frac{\sin(k_ia)}{a}$ and $k_i^2 \to \frac{2-2\cos(k_ia)}{a^2}$, (although we have set $a=1$ in our work, we include it here for the sake of being explicit) we obtain exactly our model $H_{KE}+H_{SO}$ under the condition that we identify $t = \tilde{D}/a^2$, $B = \tilde{B}/a^2$, $A=\tilde{A}/a$ and $M=\tilde{M}$. In the Hamiltonian in Eq.~\ref{hbhz} one can have\cite{rothe, lu, guigou} $\tilde{B}\sim \tilde{D}$ which translates to $B\sim t$ in terms of our parameters.  We therefore use the spin-orbit parameters in the range of Refs.~[\onlinecite{Roth,guigou}] to obtain our results. Additionally we have looked at smaller parameters for some fixed interaction variables, these are presented in the next section.

 We choose to model the interactions with effective on-site repulsion and nearest neighbor attraction, such as in the extended Hubbard model given by
\begin{equation}
V = U_0 \sum_i n_{i\uparrow}n_{i\downarrow} + V_0 \sum_{\langle i,j\rangle, \sigma,\sigma' } n_{i\sigma}n_{j\sigma'},
\end{equation}
with $U_0>0$ (repulsion) and $V_0<0$ (attraction). The motivation behind introducing an attractive $V_0$ stems from studies of a similar model without
spin-orbit coupling in the context of the cuprates.\cite{Scalapino,Onari} In those studies, it has been shown that a purely repulsive model treated in the Eliashberg formalism leads to effective off-site attraction and $d$-wave pairing on bonds.  This occurs since the pairing vertex function includes the fermionic susceptibility which has a large component close to $(\pi,\pi)$ which translates into near-neighbor attraction.  To mimic this effect in mean field we have included an attractive interaction on nearest neighbor sites.


In order to map the phase diagram of the model in Eq.~\ref{eq:FullH} we adopt a variational mean-field theory. Our
method involves obtaining a variational wave function that is a solution to an auxiliary quadratic Hamiltonian. This auxiliary Hamiltonian contains the
kinetic and spin orbit coupling parts of the Hamiltonian in Eq.~\ref{eq:FullH}.  In addition, the auxiliary Hamiltonian contains a series of quadratic terms which represent different possible orders with the order parameters as variational parameters.  These order parameters represent
all possible mean field states such as density waves, magnetism, superconductivity etc.  We have used a variety of order parameters that have appeared in similar models.  The most common density waves double the unit cell and superconductivity can occur in simple s-wave, extended s-wave and d-wave channels.  Here we present only the parameters which where found to be non-zero at some region of the phase diagram.

   The mean field ground state is found by minimizing the expectation value of
the interacting Hamiltonian ({\it i.e.} Eq.~\ref{eq:FullH}) with respect to the parameters of the variational wave function.  These parameters are
essentially the magnitudes of the various order parameters of the model.  The advantage of this method over the usual self-consistent mean-field theory is
that it does not assume {\it a priori} the dominance of any order parameter. More on its application can be found in Reference \onlinecite{VMF}.
\begin{figure}[tb]
  \setlength{\unitlength}{1mm}

   \includegraphics[scale=.5]{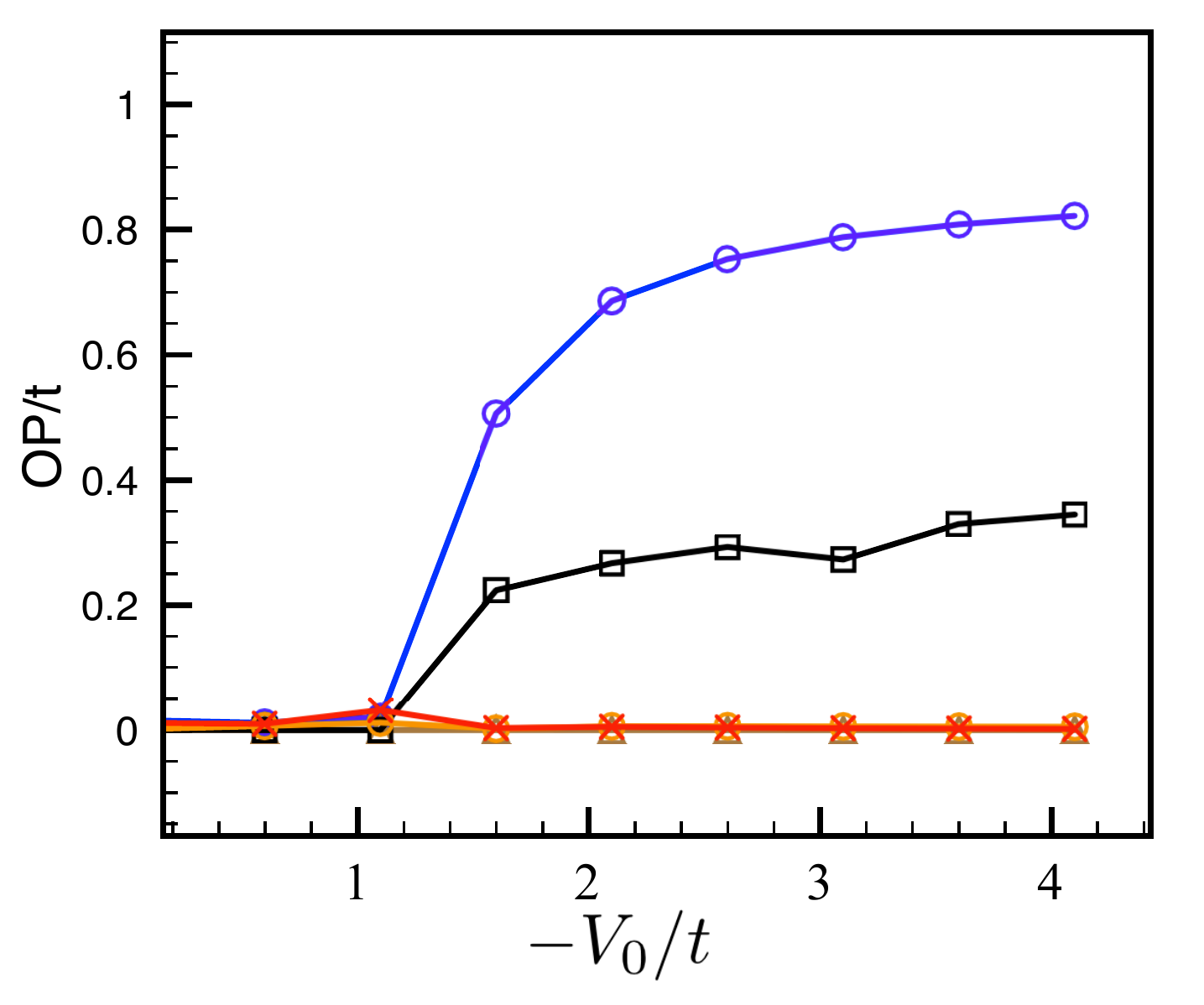}
\caption{{\small
Sample plot of order parameters. In this figure the magnitude of the order parameters is in units of $t$ and we have fixed $A=0.25t$, $B=0.5t$, $M=0.1t$ and
$U_0=2t$. Circles (blue online) are $\Delta^{(1)}$, squares (black online) are $\Delta^{(2)}$, diamonds (orange online) are $\Delta^{(3)}$, x's (red
online),  $\Delta^{(4)}$ and triangles (brown online) represent $S$. This simulation was done on a $100\times100$ square lattice.  The graph shows the
development of $d+id$ order since both $\Delta^{(1)}$ and $\Delta^{(2)}$ become non-zero at the critical coupling. For this figure we have fixed $\mu=0$.
     }
     }\label{fig:OP}
\end{figure}
\begin{figure*}[tb]
  \setlength{\unitlength}{1mm}

   \includegraphics[scale=.62]{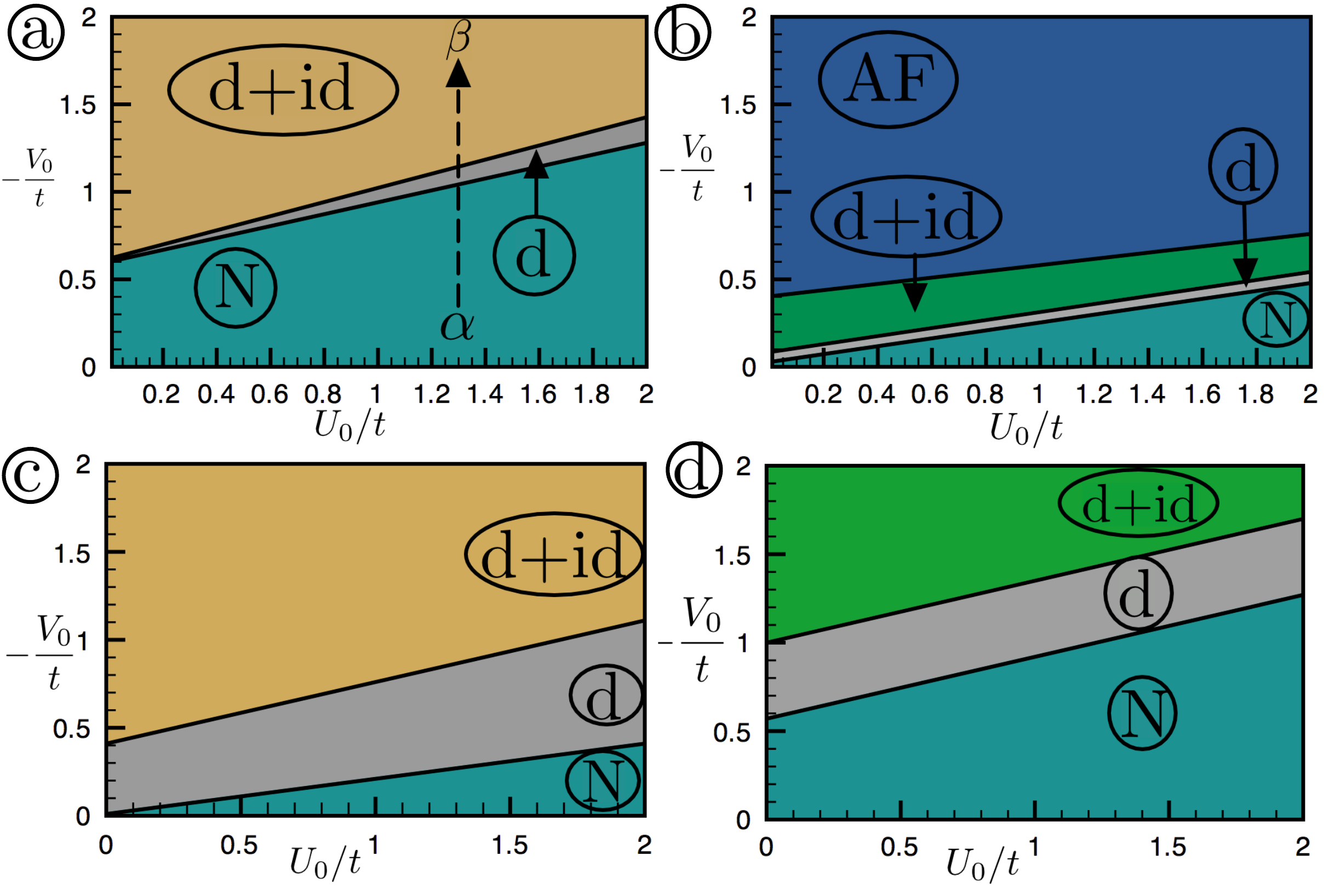}
\caption{{\small
Two dimensional slices of the phase diagram. We choose a few regions of parameter space where different phases can be observed.  The interaction parameters are scanned up to $2t$ (half the bandwidth) while the spin orbit coupling parameters and the are set at some values around the estimated values for HgTl quantum wells (Refs.[\onlinecite{rothe,guigou}]). The chemical potential has been set to zero, however, since the number of particles are not fixed and interactions are taken into account this does not imply half filling.
Specifically for panel (a) $A=0.25t, B=-0.45t$ and $M=-0.1t$, $\mu=0$. (b) $A=0.25t,B=-0.9t$ and $M=-0.05t$, $\mu=0$.  (c) $A=0.25t, B=0.5t$ and $M=0.1t$, $\mu=0$. (d) $A=0.25t, B=0.5t$ and $M=0.1t$, $\mu=.1$.
The phases are labeled by: "N" - normal, "AF" - antiferromagnetic, "d" - $d$-wave superconductor and "d+id" - a fully gapped superconductor with order parameter of the form $d_{x^2-y^2}+id_{xy}$}. The "d+id" phase in panels (a) and (c) are topologically trivial (beige online) while that of panels (b) and (d) is non-trivial (green online).}\label{fig:PD1}
\end{figure*}

To this end, we use the following auxiliary Hamiltonian
\begin{equation}
H_{\text{AUX}} = \frac{1}{4}\sum_{\kv} \Psi_{\bf k}^\dagger \Lambda_{\bf k} \Psi_{\bf k}
\end{equation}
where we have defined the $8$-spinor $\Psi_{\bf k} = (c_{{\bf k}\uparrow},c_{{\bf k}\downarrow},c_{{\bf k+Q}\uparrow},c_{{\bf
k+Q}\downarrow},c^\dagger_{-{\bf k}\uparrow},c^\dagger_{-{\bf k}\downarrow},c^\dagger_{-{\bf k-Q}\uparrow},c^\dagger_{-{\bf k-Q}\downarrow})^T$ (here ${\bf
Q}=(\pi,\pi)$)  and the matrix
\begin{equation}
\Lambda_{\bf k} = \left( \begin {array}{cc} h({\bf k}) &\hat{\Delta}({\bf k})\\ \noalign{\medskip}\hat{\Delta}({\bf k})^\dagger &-h(-{\bf k})^*\end {array}
\right),
\end{equation}
represents the Nambu space of particles and holes.  Its entries are $4\times4$ matrices:
\begin{equation}
h({\bf k}) =\left( \begin {array}{cc} \hat{\mathcal{H}}({\bf k}) &-S\sigma_z\\ \noalign{\medskip}-S\sigma_z&\hat{\mathcal{H}}({\bf k+Q})\end {array}
\right),  \ \ \hat{\mathcal{H}}({\bf k})=\epsilon_{\bf k} + \mathcal{H}_{\bf k} ,
\end{equation}
and
\begin{equation}
\hat{\Delta}({\bf k}) =\left( \begin {array}{cc} i\Delta_{\bf k}\sigma_y &0\\ \noalign{\medskip}0&-i\Delta_{\bf k+Q}\sigma_y\end {array}
\right).
\end{equation}
where $\epsilon_k=-2t(\cos{k_x}+\cos{k_y})$ is the tight binding spectrum and ${\cal H}_k$ is as defined in Eq.~\ref{SOC}.

In the above auxiliary Hamiltonian we have allowed for the possibility of antiferromagnetism (AF) through the Ne\'el order parameter $S$, as well as
several channels of superconductivity through the order parameter $\Delta_{\bf k}
=\Delta^{(1)}(\cos{k_x}-\cos{k_y})+i\Delta^{(2)}\sin{k_x}\sin{k_y}+\Delta^{(3)}(\cos{k_x}+\cos{k_y})+\Delta^{(4)}$. We can now find the variational energy
numerically for a given set of order parameters and then minimize with respect to these parameters.  This amounts to finding the mean field ground state
energy and wave function of the system.  A representative plot of these order parameters appears in Fig.~\ref{fig:OP}.

As $\Delta^{(3)}=\Delta^{(4)}=0$ throughout the plot in Fig.~\ref{fig:OP},  we conclude that this region of parameter space does not support $s$-wave nor
extended $s$-wave superconductivity. We also see that $S=0$ in Fig.~\ref{fig:OP} and so AF is also not a dominant order in this region of parameter space. The lack of
$s$- and extended $s$-wave superconductivity is a general characteristic of the phase diagram of this model, however depending on how we tune the SOC
parameters it is possible to find a state where AF (and not superconductivity) is the dominant ground state.

\section{Mean Field Phase Diagrams}

Given their various possible origins, it is difficult to estimate what the magnitude of the interaction and spin-orbit coupling parameters will be in a realistic system.  These coupling can, in principle, be determined in ab-initio calculations, however, they may vary greatly from one material to another.  We therefore explore a large portion of the $U_0-V_0$ parameter space.  In addition, other model parameters ($A, B, M$) are chosen to match known materials such as the two dimensional topological insulators for which the BHZ\cite{BHZ} model was written.

To demonstrate the differing ground states of our model we present four separate slices of the phase diagram. Fig.~\ref{fig:PD1} gives four plots in a space of the
interaction parameters $V_0$ and $U_0$; in three of the slices values of $A,B$ and $M$ are chosen so that the ground state is superconductivity, while the other
has the SOC parameters tuned so that we see a phase with an AF ground state. The $d+id$ phase in Fig.~\ref{fig:PD1} is the most interesting for our
purposes as it is fully gapped and therefore its topological invariant is well defined.  It is obtained when both $\Delta^{(1)}$ and $\Delta^{(2)}$ are
non-zero.  We view this state as having a projected superconducting order parameter whose phase winds by $6\pi$ in momentum space.  $4\pi$ of the winding
is due to its $d$-wave nature and the remaining $2\pi$ are the result of the projection on one of the spin-orbit coupled bands. We focus to this $d+id$
region of the phase diagram and investigate the topology of this phase.

\begin{figure*}[tb]
  \setlength{\unitlength}{1mm}

   \includegraphics[scale=.4]{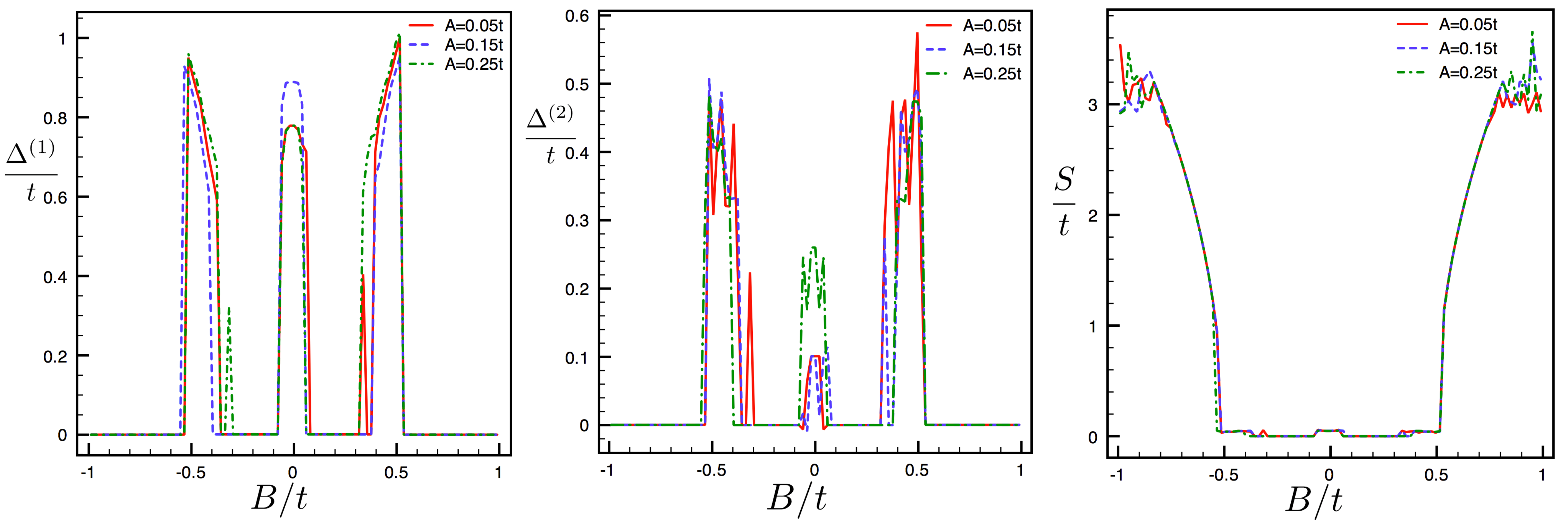}
\caption{{\small
Plot of relevant order parameters as $B$ and $A$ are changed. In this figure we have fixed $V_0=-1.8t$, $U_0=t$, and $M=-.05t$. In all three plots the solid line (red online) corresponds to $A=0.05t$, the dashed line (blue online) to $A=.15t$ and the dot-dashed line (green online) to $A=0.25t$. For convenience we have labelled these curves in the legend. 
     }
     }\label{fig:3}
\end{figure*}

Having studied four phase diagrams with numerous different ground states (from $d+id$ superconductivity to an AF) by changing the interaction strengths, we now study the dependance of various order parameters on the other parameters of the model, $A,B,M$. To do so we fix the interaction strengths $U_0$ and $V_0$ to be large enough that a phase other than the normal phase can be seen. With this in mind, Fig.~\ref{fig:3} explores the dependencies of the three order parameters $\Delta^{(1)}, \Delta^{(2)}$ and $S$ on changes in the spin-orbit coupling parameters $A$ and $B$.

Before ending this section, we make a few remarks. First, Fig.~\ref{fig:PD1} showcases a phase that is purely $d$-wave in nature. One might ask
why we are not interested in the topology of this phase; the reason is that this $d$-wave order parameter has nodes and as a result it is difficult to
properly define a topological invariant in this case.  Nevertheless, the topology of this superconductor may be an interesting topic for further studies
and could be found to be non-trivial.  On the other hand, due to nodal excitations any Majorana fermions that may be produced will not be protected against hybridizing with the low energy nodal quasiparticles.  Second, in the interest of preforming an exhaustive search for competing ground state order in this model, we have checked that neither spin
nor charge density waves give a dominant ground state contribution to the model presented here in the studied parameter regime.

\section{Discussion of Results}

Having presented several phase diagrams in the previous section this section will focus on describing and providing some physical motivation for our results. Let us begin with Fig.~\ref{fig:PD1}a where we have the three phases labeled "N", "d" and "d+id". Let us first focus on the various phases observed along a fixed value of $U_0$, for example the dashed line in the figure starting at point $\alpha$ and ending at point $\beta$. We begin in the "N" (normal) phase. Upon increasing $|V_0|$ the system undergoes a transition to a $d_{x^2-y^2}$ superconductor, what we have labeled "d". This transition to a superconducting phase can be understood by realizing that $V_0$ represents off-site attraction in our model. It is therefore understandable that a significantly strong $V_0$ should lead to pairing on nearest neighbor bonds, just the scenario in a $d_{x^2-y^2}$ superconductor. Continuing along our path we enter the "d+id" phase, a $d_{x^2-y^2}+id_{xy}$ superconductor. This transition can also be understood {\it via} the increase in $V_0$; at first $V_0$ is strong enough to induce pairing on nearest neighbors as in the "d" phase but as it is increased further it reaches a strength that is sufficient to also induce pairing on next-nearest neighbor bonds. This pairing on next-nearest neighbors is then responsible for the $id_{xy}$ term.

Increasing (decreasing) $U_0$ in Fig.~\ref{fig:PD1}a causes our $\alpha\to\beta$ contour in $U_0-V_0$ space to shift to the right (left) and increases (decreases) the strength $V_0$ required to facilitate the types of superconductivity in the "d" and "d+id" phases. This phenomenon can be understood by recalling that $U_0$ represents on site repulsion and therefore a larger $U_0$ means the probability of a site on the lattice being doubly occupied is reduced. Given that the superconducting order parameters we are interested in are proportional to $\langle c_{i+\delta, \downarrow} c_{i,\uparrow}\rangle$ ($\delta$ being a vector to nearest neighbors for  $d_{x^2-y^2}$ and next-nearest neighbors for $id_{xy}$ superconductivity) then a reduced probability for a doubly occupied lattice site should, quite roughly, lead to a decrease in this quantity as it will be less likely that site $i+\delta$ is occupied by a spin down electron and site $i$ occupied by a spin up electron. In response to this, a larger value of $V_0$ is required in order to realize the same superconducting phases as for smaller $U_0$.

Moving on the Fig.~\ref{fig:PD1}b the same general pattern is observed as in Fig.~\ref{fig:PD1}a: small values of $|V_0|$ lead to no superconductivity and as $|V_0|$ is increased  $d_{x^2-y^2}$ and $d_{x^2-y^2}+id_{xy}$ superconductivity is observed. Again in this phase diagram a stronger value of $U_0$ inhibits the superconductivity and a larger value of $|V_0|$ is required to drive pairing on nearest and next-nearest neighbors. As the general pattern is the same, all of the arguments given above for Fig.~\ref{fig:PD1}a carry over to describe the lower part of Fig.~\ref{fig:PD1}b. Despite their similarities, there are two main differences between the phase diagrams in Fig.~\ref{fig:PD1}a and Fig.~\ref{fig:PD1}b. First, the required critical values of $|V_0|$ are much lower in Fig.~\ref{fig:PD1}b and second, there is an AFM phase for large $|V_0|$ in Fig.~\ref{fig:PD1}b. The explanation of the first issue lies in the fact that we have changed both $B$ and $M$ in moving from Fig.~\ref{fig:PD1}a to Fig.~\ref{fig:PD1}b. To understand why this leads to a decrease in the critical value of $V_0$ required to develop superconductivity we must note that in these calculations we have fixed $\mu$ and so the number of particles in the system is permitted to fluctuate. For example the the phases shown in Figs. \ref{fig:PD1} and \ref{fig:3} have particle numbers varying between 1.1 and 1.3 electrons per lattice site. In general, the number of particles in the system will depend on the band structure of the system and in particular on $B$. For the phase diagram in Fig.~\ref{fig:PD1}b the system is actually closer to half filling than the system in Fig.~\ref{fig:PD1}a. Closer to half filling both the tendency to develop antiferromagnetism and $d$-wave superconductivity increase.  First, antiferromagnetism is the ground state of the Hubbard model close to half filling and therefore this tendency is not surprising.  Second, closer to half filling the system has a larger fermionic susceptibility at $(\pi,\pi)$ (as discussed in Refs.~\onlinecite{Scalapino,Onari}) and therefore the pairing vertex function in the nearest neighbor channel is enhanced.  In our case, this translates to the lower phase boundary lines in panel (b) of Fig.~\ref{fig:PD1}.

The lower two panels (panels (c) and (d)) of Fig.~\ref{fig:PD1} show slices of the phase diagram at the same values of $A,B$ and $M$ but different values of $\mu$. For small $|V_0|$ both systems start in the "N" phase and as $|V_0|$ is increased transition to the "d" phase and the "d+id" phase. The physical explanation of this behavior is the same as is given above for panel (a). The difference between the two diagrams is the critical values of $|V_0|$ as well as the ``nature" of the $d+id$-wave phase. First, the transition values in panel (c) are much lower than those in panel (d), (a) fact that, like the difference between panels (a) and (b), can be traced to the system in panel (c) being closer to half-filling. Second, as well be discussed in the next section of this paper, the $d+id$-wave phase in Fig.~\ref{fig:PD1}c is topologically trivial while that of Fig.~\ref{fig:PD1}d is topologically non-trivial. This demonstrates that we can tune $\mu$ in order to move the system across a topological phase boundary as well as the fact that a topologically non-trivial phase can be obtained for the smaller value of B, $B/t=0.5$.

Let us now turn out attention to understanding Fig.~\ref{fig:3}. The first striking feature of this figure is the relative insensitivity of any of the order parameters (and therefore phases) to changes in the value of the in plane spin-orbit coupling $A$. All three curves show small changes in the behavior of the order parameters over the range of $A$ values considered. Next, the dependance of the superconducting order parameters on $B$ consists of a peak around $B=0$ and then as $B$ is increased the order parameters drops to zero and superconductivity disappears. As $B$ is further increased we see two peaks in $\Delta^{(1)}$ and $\Delta^{(2)}$ almost evenly distributed about $B=0$. As we continue away from $B=0$ the superconducting order parameters again drop suddenly to zero and at this point the system transitions to an AFM phase as signalled by a non-zero value of $S$ in the rightmost of Fig.~\ref{fig:3}. Recalling yet again that we have held $\mu$ fixed, this crossover into an AFM phase exactly coincides with the value of $B$ for which the number of electrons in the system begins to decrease as a function of $B$.

 \section{Topological Classification}
 To study the topology of the $d+id$ region we calculate the TKNN number\cite{TKNN}(equivalent to the first Chern number) using our optimized mean-field
 wave function. This involves selecting a region in the $d+id$ phase in Fig.~\ref{fig:PD1} (or any other $d+id$ phase) and then calculating\cite{Sato1}
 \begin{equation}
 I = \frac{1}{2\pi} \int d^2k \mathcal{F}({\bf k})
 \end{equation}
 where the Berry curvature, $\mathcal{F}$, is defined using the eigenstates $\Lambda_{\kv}|\phi_n({\bf k})\rangle = E_n(\kv)|\phi_n({\bf k})\rangle$  {\it
 viz}\cite{Berry}
 \begin{equation}
  \mathcal{F}({\bf k}) = i\sum_{n}' \sum_{m\ne n} \epsilon^{ij}\left[\frac{\langle \phi_n|\frac{\partial \Lambda_{\kv}}{\partial k_i}|\phi_m\rangle\langle
  \phi_m|\frac{\partial \Lambda_{\kv}}{\partial k_j}|\phi_n\rangle}{(E_n-E_m)^2}\right],
 \end{equation}
where, for the sake of brevity, we have dropped the functional dependence on ${\bf k}$,  the primed sum is a sum over filled bands, the $\epsilon^{ij}$
tensor has the values $\epsilon^{1,2}=-\epsilon^{2,1}=1$ and $\epsilon^{i,i}=0$ and summation over the repeated indices $i$ and $j$ is implied.

\begin{figure}[tb]
  \setlength{\unitlength}{1mm}

   \includegraphics[scale=.52]{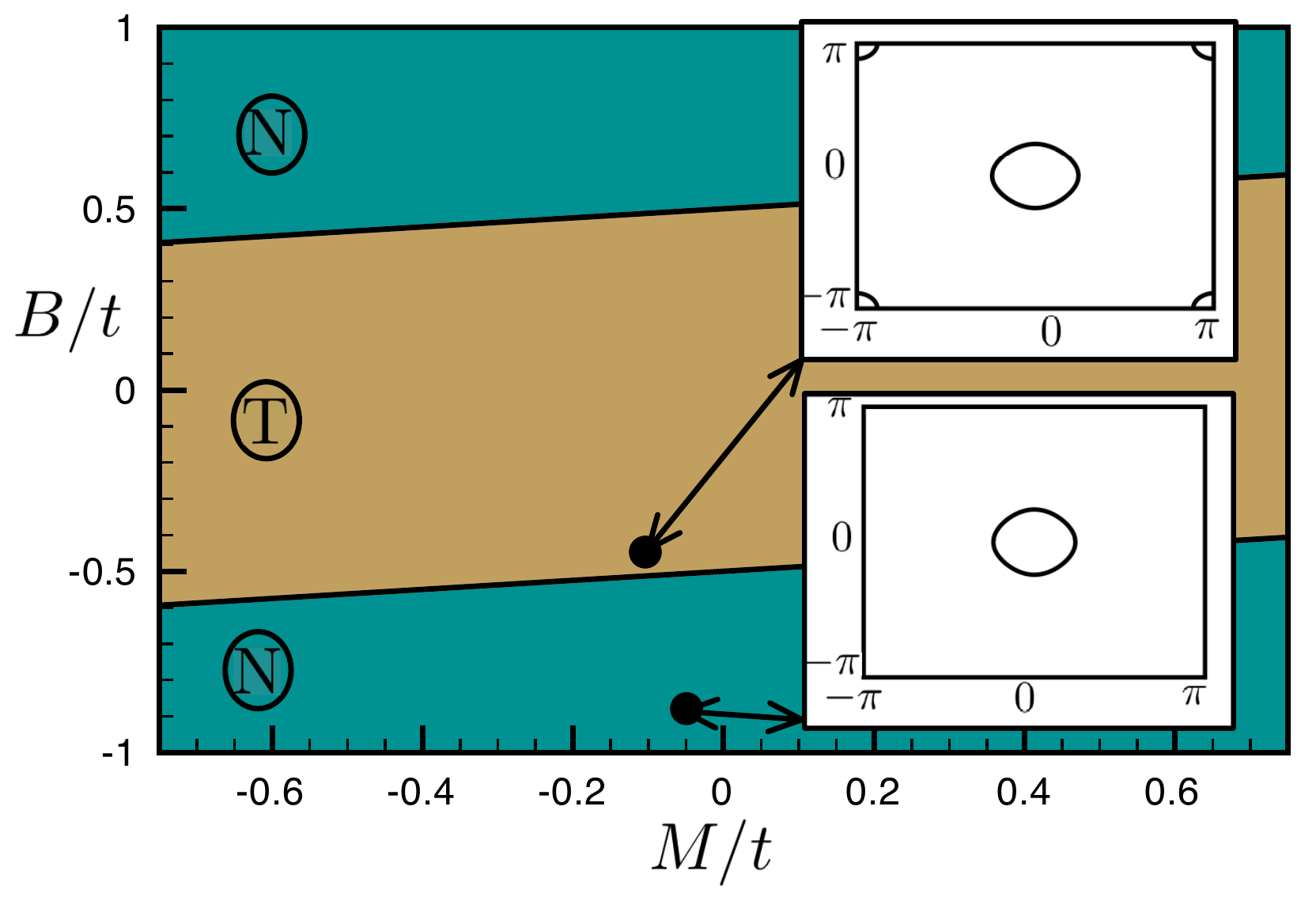}
\caption{{\small
Topology of the $d+id$ ground state phase. In this figure we have set $A=0.25t$ and $\mu=0$. The topologically trivial phase is labeled T (beige online) while the
non-trivial phase is labeled "N"(teal online). The insets show an example of the Fermi surface (in the first Brillouin zone) for each phase before
interactions are turned on. The lower inset, corresponding to the $d+id$ state of Fig.~\ref{fig:PD1}b, shows a single Fermi surface for the topologically non-trivial
phase while the upper inset (the $d+id$ state in Fig.~\ref{fig:PD1}a) shows two Fermi surfaces in the topologically trivial region. Note that as we tune $\mu$ this diagram maintains this same general behavior the only difference being that the absolute value of the $B$-intercept of the two boundary lines increases (decreases) for decreasing (increasing) $\mu$.
     }}\label{fig:top}
\end{figure}

 By calculating this invariant we can classify the topology of the $d+id$ region as either trivial (regions for which we find $I=0$) or non-trivial
 (regions for which $I=1$). Our results are summarized in Fig.~\ref{fig:top}. As the topology of the system is intimately related to the number of Fermi surfaces
 before interactions are turned on\cite{FuBerg}, Fig.~\ref{fig:top} also shows a sample of the Fermi surface in each topological region.

Fig.~\ref{fig:top} displays one of our main results; our model has regions of $d+id$ topological superconductivity. As an example, Fig.~\ref{fig:top}
shows that the $d+id$  state in Fig.\ref{fig:PD1}a is topologically trivial while the like in Fig.~\ref{fig:PD1}b is topological. Further, we see from the figure some devices
might exist in the topologically non-trivial region at some set value of $M$. It may then be possible to move the system into a topological phase by
changing $M$ via either an applied field or proximity to a magnetic layer. In this way, our results suggest that properly applying a Zeeman field ({\it i.e.} tuning $M$) to spin-orbit coupled superconductors may result in the transition of an ordinary superconductor to a topological one.

The reader should also note the strengths of the parameters $B$ and $M$ required to obtain non-trivial topology. We see that for the range of parameters shown in Fig. \ref{fig:top}, a value of $B\gtrsim \pm  0.5t$ is required for non-trivial topology. In order to find a non-trivial topological state for $B\simeq 0.0$, a very large value of $|M|>4t$ is required (not shown in Fig.  \ref{fig:top}). This might suggest that a quantum well type system may be the most suitable for realizing the topological superconductor as for these systems the typical values of $B$ are large. 

In a real material, it is not possible to tune the interaction.  However, this is routinely done in cold atoms.  It has been demonstrated recently that
spin-orbit coupling may be simulated in cold atoms\cite{Lin}.  This may lead the way for simulating topological insulators\cite{Mei,Goldman,Essin}.  Our
work suggests that if the spin-orbit coupling and the interactions are tuned correctly, a topological superconductor may be simulated as well.


\section{Conclusions}
In summary, we have proposed a model of interacting, spin-orbit coupled, Zeeman split electrons on a square lattice. We have shown
that in some regions of parameter space the ground state of the proposed model is either a $d+id$ superconductor or an antiferromagnet. Narrowing our focus to
the $d+id$, region we have shown that our system supports phases of non-trivial topology. The topological regions in our phase diagrams exhibit
superconductivity with $6\pi$ winding of its order parameter phase in the Brillouin zone.  In the same way that a $p+ip$ superconductor may
support the existence of Majorana fermions in vortex cores or on edges, this superconductor will support their existence as well.  As our model considers
superconductivity driven by interactions rather than the proximity effect, it may serve as a possible simplification for device design. Finally, our work
supplies strong evidence that Majorana fermions might be realized in certain spin-orbit coupled superconductors under the proper application of a Zeeman
field. The work presented here provides an initial study of a model that is very rich in the sense that it could be used to describe various different scenarios. Future work in the direction of the results presented here would focus on finding a specific material that falls in the topological superconducting phase we have found.

\section{Acknowledgements}
The authors would like to thank J. Alicea, E. Berg, B. A. Bernevig, A. A. Clerk, G. Gervais and S. Sachdev for useful
conversations. This work was supported by Natural Sciences and Engineering Research Council of Canada (AF, TPB) as well as the McGill Tomlinson Fellowship program (AF). A number of the numerical calculations in this work were preformed using CLUMEQ supercomputing resources.

\bibliographystyle{apsrev}
\bibliography{TopoSC}
\end{document}